\documentstyle[a4,12pt,epsfig]{article}
\def\beq{\begin{equation}}
\def\eeq{\end{equation}}
\def\bea{\begin{eqnarray}}  \def\eea{\end{eqnarray}}
\def\lsim{\raise0.3ex\hbox{$<$\kern-0.75em\raise-1.1ex\hbox{$\sim$}}}
\def\gsim{\raise0.3ex\hbox{$>$\kern-0.75em\raise-1.1ex\hbox{$\sim$}}}
\def\1{{\rm 1\mskip-4.5mu l} }
\newcommand{\noi}{\noindent}
\begin{document}

\begin{center}
\vspace*{1 truecm}
{\Large \bf Charmonium Suppression in Lead-Lead Collisions:} \\
{\Large \bf  Is There a Break in the $J/\psi$ Cross-Section?} \\[8mm]
{\bf N. Armesto}\par
{\it II. Institut f\"ur Theoretische Physik, Universit\"at Hamburg,} \\
{\it Luruper Chaussee 149, D-22761 Hamburg, Germany} \\
[5mm]
{\bf A. Capella and E. G. Ferreiro}\par
{\it Laboratoire de Physique Th\'eorique et Hautes
Energies,}
\\ {\it Universit\'e
de Paris-Sud, B\^atiment 211, F-91405 Orsay Cedex, France}
\end{center}

\vskip 1 truecm
\begin{abstract}
In the framework of a model based on nuclear absorption plus comover
interaction, we compute the
$E_{T}$ distribution of the $J/\psi$ in $PbPb$ collisions at SPS and compare it
with available
NA50 data.
Our
analysis suggests
that
the existence of new physics (deconfinement phase transition) in the region
$E_{T} \ \lsim \ 100$ GeV is unlikely
and that signals of new physics should rather be
searched in the
region $E_T \ \gsim \ 100$ GeV. The $E_{T}$ dependence of the $J/\psi$
transverse momentum
has been computed. At large $E_T$ it turns out to be much flatter in the
comover
approach than in a phase transition framework. Estimates of the $J/\psi$
suppression at RHIC and
LHC energies are also given.

\end{abstract}

\vfill

\noindent LPTHE Orsay 98/27  \par
\noindent DESY 98-086  \par
\noindent hep-ph/9807258  \par
\noindent September 1998

\newpage
\pagestyle{plain}
\section{Introduction}
\hspace*{\parindent}
The 1995 data from the NA50 Collaboration \cite{1r} show an anomalous $J/\psi$
suppression, i.e. a
suppression larger than the one expected in a nuclear absorption model. This
model describes the
$J/\psi$ suppression both in proton-nucleus interactions and in nucleus-nucleus
interactions with
a light projectile \cite{1r}. Following the original proposal of \cite{1br},
the anomalous
$J/\psi$ suppression has been interpreted as a signal of a deconfining phase
transition
\cite{2r,3r,4r,5r}. However, an explanation by a more conventional mechanism,
namely, the
interaction of the $J/\psi$ (or the $c\overline{c}$ pair) with comovers, is
also possible
\cite{6r,7r,8r,9r}.

A very spectacular feature of the 1996 data by the same Collaboration
\cite{10r}, is the presence
of a break in the ratio $R(E_{T})$ of $J/\psi$ over Drell-Yan ($DY$)
cross-sections at $E_{T}
\sim$ 55 GeV. It has been argued \cite{11r,5r} that this break is a signal of
deconfinement --
although there is no general consensus on this point \cite{12r,2r,3r,4r,5r}. It
is, however, fully
recognized that a break in the $J/\psi$ cross-section would rule out any
conventional model, such
as the one based on comover interaction. (Other approaches to $J/\psi$
suppression can be found in
\cite{11br}.)

At present, the evidence for this break is weakened by the presence of
fluctuations in the ratio
$R(E_{T})$ (see Eq. (\ref{13e}) for a precise definition) at large values of
$E_{T}$ \cite{10r},
which are generally regarded as spurious. Also, it is necessary to assess
whether this break, if
confirmed, is due to a genuine break in the $J/\psi$ cross-section or rather to
fluctuations in
the $J/\psi$ and $DY$ ones. Of course, a definitive answer to these questions
can only come from  data.
However, in view of the interest of the subject, it is important to examine the
available data in
a theoretical framework in order to gain some insight on these questions, while
waiting for a
complete analysis of the 1996 NA50 data, and above all, for the 1998 results.

The aim of the present work is to perform such an analysis in the framework of
a model based on
nuclear absorption plus comover interaction. This work is a continuation of the
one in Ref.
\cite{8r}. We use the same formalism and the same values of the parameters
which were determined
in \cite{8r} from the best fit to the ratio $R(E_{T})$ in $pA$, $SU$ and $PbPb$
collisions. The
plan of the paper is the following. In Section 2 we describe the model. In
Section 3 we compute the
$E_{T}$ distributions of minimum bias, $DY$ and $J/\psi$, and compare them with
available data. In
Section 4 we compute the ratio $R(E_{T})$ and compare it with the NA50 data. In
Section 5 we
compute the $E_{T}$ dependence of the average $p_{T}^2$ of the $J/\psi$ and
compare it with recent
NA50 data. Section 6 contains our conclusions and prospects at higher energies.

\section{The model} \hspace*{\parindent} Our model is formulated in a
conventional framework
\cite{6r,7r,8r,9r} based on two different mechanisms of $J/\psi$ suppression:
nuclear absorption
of the pre-resonant $c\overline{c}$ pair with nucleons of the colliding nuclei
and absorption by
comoving partons or hadrons produced in the collision. For completeness we
recall its main
ingredients: \par \vskip 5 truemm

\noindent \underbar{\bf Nuclear absorption:} In nucleus-nucleus collisions, the
survival
probability of the $J/\psi$ at impact pa\-ra\-me\-ter $b$ and transverse
position $s$ is given by
\cite{13r,8r} \beq S^{abs}(b, s) = {[1 - \exp (- A \ T_A(s) \ \sigma_{abs}) ]
[1 - \exp (- B\
T_B(b-s) \ \sigma_{abs})] \over \sigma_{abs}^2 \ AB \ T_A(s) \ T_B(b-s)} \ \ .
\label{1e} \eeq
Here $T_{A(B)}(b)=\int_{-\infty}^{+\infty} dz \ \rho_{A(B)}(b,z)$ are the
nuclear profile
functions normalized to unity. The nuclear densities $\rho_{A(B)}(b,z)$ are
determined from a
3-parameter Fermi distribution with parameters given in Ref. \cite{14r}. (In
Ref. \cite{8r} a
different parametrization of the nuclear density is used; this introduces
differences in $R(E_T)$
of less than 4 \% in $PbPb$ collisions.) For the absorptive cross-section we
take $\sigma_{abs} =
6.7\div 7.3 $ mb, consistent with a fit to the proton-nucleus data \cite{4r}.
Note that $S^{abs} =
1$ for $\sigma_{abs} = 0$. \par \vskip 5 truemm

\noindent \underbar{\bf Absorption by comovers.} This absorption is due to the
interaction of the
$c\overline{c}$ pair (or of the $J/\psi$ itself) in the dense medium produced
in a nucleus-nucleus
collision -- which results in the production of a $D\overline{D}$ pair. The
$J/\psi$ survival
probability is given by \cite{4r,8r} \beq S^{co}(b, s) = \exp \left [-
\sigma_{co} \ N_y^{co}(b,
s)\   \ln \left ( {N_y^{co}(b, s) \over N_f} \right )\ \theta (N_y^{co}(b, s) -
N_f)  \right ] \ \
. \label{2e} \eeq Here $N_y^{co}(b, s)$ is the initial density of comovers per
unit transverse
area $d^2s$ and per unit rapidity at impact parameter $b$, and $N_f$ is the
corresponding
freeze-out density. In order to have a smooth onset of the comovers, it is
natural to take for
$N_f$ the density of hadrons per unit rapidity in a $pp$ collision, i.e. $N_f =
[3/(\pi R_p^2)]\
dN^-/dy|_{y^* = 0} = 1.15 \ {\rm fm}^{-2}$. This coincides with the value
introduced in Ref.
\cite{4r}. With this choice of $N_f$, the $\theta$-function in Eq. (\ref{2e})
is numerically
irrelevant. The effect of the comovers in $pA$ turns out to be negligibly
small. $\sigma_{co}$ is
the comover cross-section properly averaged over the momenta of the colliding
particles (the
relative velocity of the latter is included in its definition). The logarithmic
factor in Eq.
(\ref{2e}) is the result of an integration in the proper time $\tau$ from the
initial time to
freeze-out time. (One assumes \cite{15r,16r} a decrease of densities with
proper time in
$1/\tau$.) A large contribution to this integral comes from the few first fm/c
after the collision
-- where the system is in a pre-hadronic stage. (In this respect, see the last
paper of Ref.
\cite{9r}.) Actually, Brodsky and Mueller \cite{17r} introduced the comover
interaction as a
coalescence phenomenon at the partonic level. In view of that, there is no
precise connection
between $\sigma_{co}$ and the physical $J/\psi - \pi$ or $J/\psi - N$
cross-section, and
$\sigma_{co}$ has to be considered as a free parameter. We take $\sigma_{co} =
0.6$ mb \cite{8r}.
\par \vskip 5 truemm

\noindent \underbar{\bf Cross-sections:} The $J/\psi$ production cross-section
in nuclear
collisions is given by \beq \label{3e} \sigma_{AB}^{\psi}(b) =
{\sigma_{pp}^{\psi} \over
\sigma_{pp}} \int d^2s \  m(b,s) \ S^{abs}(b,s) \ S^{co}(b, s)\ \ , \eeq where
\beq \label{4e}
m(b, s) = AB \  \sigma_{pp} \ T_A(s) \ T_B(b - s) \ \ . \eeq We take
$\sigma_{pp} = 30$ mb. With
the definition (\ref{3e}), the Drell-Yan cross-section (obtained from
(\ref{3e}) with
$\sigma_{abs} = \sigma_{co} = 0$) is proportional to $AB$.

The cross-section for minimum bias ($MB$) events is given by \beq \label{5e}
\sigma_{AB} (b) = 1 -
\exp [- \sigma_{pp} \ AB \ T_{AB}(b)]\ \ , \eeq with $T_{AB}(b) = \int d^2s \
T_A(s) T_B(b - s)$.

In order to compute these cross-sections we need to know the comover density
$N_y^{co}(b, s)$ in
the NA50 dimuon spectrometer. Moreover, comparison with experiment requires to
compute the above
cross-section at a given transverse energy $E_{T}$ -- measured in the NA50
calorimeter. This
requires the knowledge of the $E_T - b$ correlation function $P(E_T, b)$. In
the following we
proceed to calculate these two quantities. \par \vskip 5 truemm

\noindent \underbar{\bf Density of comovers:} It is commonly assumed in the
literature that the
density of comovers is proportional to that of participating (or wounded)
nucleons \cite{4r,18r}.
This is the so-called Wounded Nucleon Model (WNM; for a review see
\cite{19r}). In asymmetric
systems and, in particular, in $pA$ collisions, this model provides a
reasonable description of
the data but only for the average multiplicity -- or at negative rapidities,
close to the maximum of
the rapidity distribution. For symmetric $AA$ collisions, the model seems to be
valid in a broader
rapidity range. (This can be understood from the arguments in Ref. \cite{21r};
see p. 26.)
However, for central $PbPb$ collisions (and also for other central
nucleus-nucleus collisions at
SPS) there is experimental evidence of a violation of this scaling law at
mid-rapidities
\cite{20r,20br}. Moreover, models such as the Dual Parton Model (DPM)
\cite{21r}, in which
unitarity is fully implemented, contain an extra term proportional to the
average number of
collisions. This term is small at present energies but its relative size
increases with energy.
Moreover, it contributes mostly at mid-rapidities. The origin of this term is
the following. In DPM
one has both baryonic strings of type diquark-quark and bosonic ones of type
$q$-$\overline{q}$.
The latter contribute mainly at mid-rapidities. Since the number of diquarks
available is equal to
the number of participating nucleons, the number of baryonic strings is equal
to the number of
par\-ti\-ci\-pants. On the other hand, the total number of strings is
proportional to the number
of collisions. Therefore, the number of $q$-$\overline{q}$ strings increases,
with increasing
centrality, much faster than the number of participants. The WNM is obtained
from DPM by
neglecting the contribution of the $q$-$\overline{q}$ strings.

In the following calculations we will use the density of comovers given by DPM.
We will also
discuss how the $J/\psi$ suppression is modified when using a density of
comovers proportional to
the number of participants.

In DPM, $N_y^{co}(b, s)$ is given by \cite{8r,21r} \bea && N_y^{co}(b, s) =
\left [ N_1 \ m_A(b,
s) + N_2 \ m_B(b, b - s) + N_3 \ m(b,s) \right ]\theta ( m_B(b,b - s) -
m_A(b,s)) \nonumber \\
 &+& [N^\prime_1 \ m_A(b, s) + N^\prime_2 \ m_B(b,b - s) + N^\prime_3 \ m(b, s)
] \theta \left (
m_A(b, s) - m_B(b, b - s) \right ) \ \ . \label{6e} \eea Here $m$ is given by
Eq. (\ref{4e}) and
$m_A$, $m_B$ are the well known geometric factors \cite{22r,16r} \beq
m_{A(B)}(b, s) = A(B) \
T_{A(B)} (s) \left [ 1 - \exp \left ( - \sigma_{pp} \ B(A) \ T_{B(A)}(b - s)
\right ) \right ] \ \
. \label{7e} \eeq The coefficients $N_i$ and $N^\prime_i$ are obtained in DPM
by convoluting
momentum distribution functions and fragmentation functions \cite{21r}. Their
values (per unit
rapidity) for the rapidity window and energies of the NA38 and NA50 experiments
are given in Table
1 of Ref. \cite{8r}. The rapidity density of hadrons is given by

\beq {dN^{co} \over dy} = {1
\over \sigma_{AB}} \int d^2b \int d^2s \ N_y^{co}(b, s) \ \ \ ,  \label{8bis}
\eeq with
$\sigma_{AB} = \int d^2b\ \sigma_{AB}(b)$. Note that at fixed $b$ in the range
of interest,
$\sigma_{AB}(b) \simeq 1$.

The obtained densities of negative hadrons at $y^* = 0$ for $pp$, $SS$, $SAu$
and $PbPb$ are
compared in Table 2 of Ref. \cite{8r} with available data, using in each case
the centrality
criteria (in percentage of total events) given by the experimentalists. In Fig.
1 we compare the
predictions of both the DPM and the WNM with the NA49 data \cite{20r} for the
rapidity
distribution of negative particles in central $PbPb$ collisions at 158 AGeV/c.
\par \vskip 5 truemm

\noindent \underbar{\bf $E_T - b$ correlation:} The experimental results are
given as a function
of $E_T$. This is the total transverse energy of neutrals measured by the NA50
calorimeter in the
rapidity window $- 1.8 < y^* < - 0.6$. The correspondence between average
values of $b$ and $E_T$
is given by the proportionality between $E_T$ and multiplicity:
\beq \label{8e} E_T(b) = q \
N_y^{co}(b)\ \ , \eeq
where $N_y^{co}(b) = \int_{-1.8}^{-0.6} dy \int d^2s \ N_y^{co}(b, s)$, with
$N_y^{co}(b, s)$ given by Eq. (\ref{6e}). The parameter $q$ is closely
connected to the average
transverse energy per particle. However, it contains extra factors due to the
fact that $N_y^{co}$
corresponds to the multiplicity of negatives whereas $E_T$ is the transverse
energy of neutrals.
Moreover, a calibration factor of the NA50 calorimeter (which has an estimated
systematic error of
about 40 \%) is also included in $q$.

A precise determination of $q$ comes from the measured correlation between
$E_T$ and $E_{ZDC}$ --
the energy measured at the zero-degree calorimeter. The latter is defined as
\beq \label{9e}
E_{ZDC}(b) = \left [ A - m_A(b) \right ] \ E_{in}\ \ , \eeq
 where $m_A(b) = \int d^2s \ m_A(b,
s)$, i.e. the average number of participants of $A$ at fixed impact parameter,
and $AE_{in}$
is the beam energy ($E_{in}=158$ GeV/c).
A fit to the experimental $E_T - E_{ZDC}$ correlation using
Eqs.
(\ref{8e}) and (\ref{9e}) allows a precise determination of $q$. From the NA50
data \cite{10r}
we obtain $q = 0.78$ GeV. It follows from (\ref{8e}) and (\ref{9e}) that with
the WNM
ansatz \cite{4r}: $E_T(b) = 0.4 \ [m_A(b)+m_B(b)]$ GeV, the $E_T - E_{ZDC}$
correlation is a straight line. Experimentally, it is indeed found to be close
to a straight line
but shows a clear concavity. In DPM this correlation has a concavity due to the
contribution of the
$q$-$\overline{q}$ strings. However, in the acceptance region of the NA50 $E_T$
calorimeter, the
contribution of the $q$-$\overline{q}$ strings is rather small (see Fig. 1) and
the concavity is
also small. Actually, DPM describes well the data in the upper half of the
$E_T$ region but falls
too fast at low $E_T$ -- while the WNM describes the data better in the low
$E_T$ region (see Fig.
2). One could think that the difference between the two correlation functions
is too small to
have any significant effect on the shape of the $E_T$ distributions. It turns
out that this is not
the case and, therefore, a more accurate description of the $E_T - E_{ZDC}$
correlation is needed.
\par

The failure of the DPM at low $E_T$ can be attributed to the effect of the
intra-nuclear cascade,
which is not included in (\ref{6e}). This well known phenomenon consists in the
production of
extra particles in the fragmentation regions of the two colliding nuclei due to
the rescattering of
slow secondaries (in the rest frames of the two nuclei) with spectator
nucleons. Obviously this
effect has to vanish for central collisions when no spectator nucleons are
left. It is also absent
at mid-rapidities. However, the rapidity region of the $E_T$ calorimeter $- 1.8
< y^* < 0.6$ is
affected by the intra-nuclear cascade (which is known to have an extension of
about 1.5
rapidity units). In order to incorporate the intra-nuclear cascade in a
phenomenological way, we
replace Eq. (\ref{8e}) by

$$E_T(b) = q\ N_y^{co}(b) + k \ E_{ZDC}(b) \quad . \eqno(9')$$

\noi With the values of the parameters we use, $q = 0.78$ GeV and $k = 1/4000$,
the relative contribution of
the second term in ($9^\prime$) is comparatively small (about 30 \% for a very
peripheral collision with
$E_{ZDC} = 30 000$ GeV and less than 2 \% for $E_{ZDC} \ \lsim \ 10000$ GeV).
The only drawback of this extra term is that it
does not vanish at $E_{ZDC} =
E_{ZDC}^{MAX} = AE_{in}$. However, this can be easily cured by replacing Eq.
($9^\prime$) by

$$E_T(b) = q \ N_y^{co}(b) + 0.95 \left ( {E_{ZDC}(b) \over 4000} \right
)^{1.2} \left (
{E_{ZDC}^{MAX} - E_{ZDC}(b) \over E_{ZDC}^{MAX}} \right )^{0.2} \quad
.\eqno(9'')$$

\noi The corresponding $E_T - E_{ZDC}$ correlation, shown in Fig. 2
(full line), is practically identical to
the one obtained
from  ($9^\prime$)
for $E_T < 30 000$ GeV and gives
an excellent description of the experimental data \cite{10r}.
Moreover, both correlations lead to the same
$E_T$ distributions
for $J/\psi$ and $DY$ in the region $E_T \ \gsim \ 15$ GeV, where data are
available. Eq. ($9^{\prime \prime}$) will be
used in all DPM calculations. \par

In order to obtain the $E_T - b$ correlation, and not only the relation between
the
average values of these two quantities, we have to determine the $E_T$
distributions
at a given $b$. A good description of the experimental $E_T$ distributions is
obtained
\cite{18r,4r,24r} using a Gaussian distribution at fixed impact parameter,
with
squared dispersion $D^2(b) \equiv
\langle [N_y^{co}(b)]^2\rangle  - \langle N_y^{co}(b)\rangle ^2 = a\langle
N_y^{co}(b)\rangle $, i.e.
\beq
\label{10e}
P(E_T , b) = {1 \over \sqrt{2 \pi q^2 a\overline{N}_y(b)}}
\exp \left [ - {[ E_T -
q\overline{N}_y(b)]^2 \over 2q^2a\overline{N}_y(b)} \right ]\ \ ,
\eeq
where $\overline{N}_y(b) = E_T(b)/q$, with $E_T(b)$ given by Eq.
($9^{\prime \prime}$), and $a$ is a free
parameter (see Section 3).

\section{${\bf E}_{\bf T}$ distributions} \hspace*{\parindent}
The $E_T$ distributions of $J/\psi$, $DY$ and Minimum Bias $(MB)$ are obtained
by folding the
cor\-res\-pon\-ding cross-sections at fixed $b$ (Eqs. (\ref{3e}) and
(\ref{5e}))
with
the $E_T - b$ correlation function:
\beq
\label{11e}
{d\sigma^{\psi} \over dE_T} = \int d^2b \ \sigma_{AB}^{\psi}(b) \ P(E_T, b)\
\ ,
\eeq
\beq
\label{12e}
{d\sigma^{MB} \over dE_T} = \int d^2b \ \sigma_{AB}(b) \ P(E_T, b) \ \ .
\eeq
The corresponding expression for $DY$ is obtained from (\ref{11e}) with
$\sigma_{abs} = \sigma_{co} = 0$.

The most precise determination of the parameter $a$ is
obtained from a fit of
(the tail of) the $MB$
$E_T$ distribution. Using the 1995 data of Ref. \cite{24r}
we obtain $a =
0.73$. This value will be used in all
DPM calculations. With the WNM we use the parameters in Ref.
\cite{24r}: $q=0.4$ GeV
and $a=1.43$. Note that the product $aq$ is the same in both cases. (It
turns out that
the ratio of $J/\psi$ over $DY$ is very insensitive to the value of $a$.)
\par

The comparison of $d\sigma^{DY}/dE_T$ with the 1995 data \cite{24r} is shown in
Fig. 3. The
agreement is satisfactory but the error bars are quite large. Also shown is the
distribution
obtained using the WNM. This correlation has a stronger increase with increasing
$E_T$ -- but both are
consistent with the data within errors. \par

The comparison with the 1996 $E_T$ distribution is shown in Fig. 4. Again the
(statistical) error bars are quite
large. Moreover, there is a significant disagreement both with DPM and WNM at
$E_T \sim 135$ GeV
which was not present when comparing with the 1995 data. There is also a
significant difference in
shape between DPM and WNM. Note that the only ingredients in the calculation
are the $b$
dependence of the $DY$, $ABT_{AB}(b)$, which is common to all models, plus the
$E_T - b$ or $E_T -
E_{ZDC}$ correlation. Thus all models which reproduce the latter correlation
should lead to
the same $DY$ distribution. Since the DPM (with Eqs. ($9^\prime$)
or ($9^{\prime \prime}$) and
(\ref{9e})) gives an
excellent description of the latter, the full curve in Fig. 4 should be
regarded as the
theoretical $DY$ distribution -- which can be used as a reference when
considering the  $J/\psi$
one. Any significant discrepancy with this distribution, such as the one
occurring at large $E_T$,
should be regarded as a possible experimental inconsistency between the
measured $DY$ $E_T$
distribution and the $E_T - E_{ZDC}$ correlation. \par

We turn next to the $E_T$ distribution of the $J/\psi$. We have computed
it using $\sigma_{abs}=6.7$ mb, $\sigma_{co}=0.6$ mb. The result of our
calculation is compared with the 1995 data \cite{24r} of the NA50 Collaboration
in Fig. 5. The agreement between theory and experiment is reasonable. However
the data seem to
decrease slightly faster than the theoretical curve. Note that these data show
no break in the
$J/\psi$ cross-section at any value of $E_T$. Fig. 5 shows also the $E_T$
distribution obtained with nuclear absorption alone ($\sigma_{abs} = 7.3$ mb)
both for DPM (Eq.
($9^{\prime \prime}$))
and for the WNM. We see that the shape of the $E_T$ distribution is very
sensitive to the
effect of the comovers. Thus, a slightly steeper decrease of the $J/\psi$
cross-section, if
confirmed, could possibly be obtained with a small increase in the absorption
parameters. (The
constraint on these parameters coming from the $SU$ data is now significantly
smaller due to an
increase by a factor 2.8 of the statistical errors; see Section 6.) Note also
that, with nuclear
absorption alone, the WNM has a faster increase with $E_T$ than the DPM.
Therefore the extra
$J/\psi$ suppression required in order to reproduce a given shape of the
$J/\psi$ distribution,
must be considerably stronger in the WNM than in DPM. This is even more clearly
seen in Fig. 6 where we
compare the theoretical predictions with the $J/\psi$ $E_T$ distribution from
the 1996 data in a
linear scale. In this comparison we observe some deviations at $E_T \ \gsim \
100$ GeV.
This region should be studied with great care in
the 1998 high
statistics run. In our opinion, this is a most interesting region to look for
eventual signs of
new physics, i.e. for the onset of a truly anomalous suppression at $E_T \ \gsim
\ 100$ GeV. On the
contrary, in the region $E_T < 100$ GeV (where the break in the ratio $R(E_T)$
occurs), 
% CAMBIO
%we have a
%reasonable agreement 
there is no strong disagreement 
between theory and experiment.
%There are
%small fluctuations of the data around the theoretical curve
%-- with a small
%minimum at $E_T \simeq 55$ GeV, corresponding to the break in $R(E_T)$, and
%small maxima around
%40 and 80 GeV. 
However, there is no 
perfect agreement either and, therefore, it is not possible to draw a clear 
conclusion at present. In particular a sudden drop of the $J/\psi$
cross-section at $E_T \simeq 55$ GeV has been claimed \cite{10r}. 
Even if further data show that this drop is statistically significant, our
analysis suggests that it cannot be easily attributed to a sudden increase of 
$J/\psi$ suppression due to deconfinement. Indeed, in the next four $E_T$
bins the measured $J/\psi$
cross-section is consistent with the predictions of a model which does not 
have deconfinement.

\section{$J/\psi$ over $DY$ ratio}
\hspace*{\parindent}
The $J/\psi$ suppression is described by the ratio $R(E_T)$ of $J/\psi$ and
$DY$ cross-sections in
different $E_T$ bins. The advantage of taking this ratio is that systematic
errors common to both
systems do not appear in this ratio. The inconvenient, however, is that the
results are sensitive
to the shape of the $DY$ $E_T$ distribution. In our opinion, it is of utmost
importance to have
good data on the $E_T$ distribution of the $J/\psi$ -- as illustrated by the
analysis of the
previous Section. The ratio $R(E_T)$ is given by
\beq \label{13e} R(E_T) = {\int d^2b \
\sigma_{AB}^{\psi}(b) \ P(E_T,b) \over \int d^2b \  \sigma_{AB}^{DY} \
P(E_T,b)} \ \ . \eeq
This ratio has been calculated, within the present model, in Ref. \cite{8r},
where the values of the
parameters (the same ones used here, including the absolute normalization but
excepting the value
of $a$ which, as discussed in the previous Section, has practically no effect
on $R(E_T)$) were
determined from the best fit to $R(E_T)$ in $pA$, $SU$ and $PbPb$ collisions.
At that time,
however, the 1996 data were not available. The comparison of the model results
with both the 1995
and 1996 data \cite{10r} is shown in Fig. 7.

The model reproduces the qualitative behavior of the ratio $R$. However, there
are disagreements
at a quantitative level. The overall suppression, both from the 1995 and the
1996 data, is somewhat
larger than the theoretical one. More important, the 1996 data seem to show a
break at $E_T \sim 55$ GeV
which is not present in the model calculation. Here, several comments are in
order. First, as seen
in Fig. 7, the experimental data for the first  $E_T$ bin is higher than the
one obtained with
nuclear absorption alone (with a normalization extracted from a fit to $pA$ and
$SU$ data
\cite{1r,8r}). This is difficult to explain in any model. Second, the relevance
of the break at
$E_T \sim 55$ GeV is weakened by the existence of fluctuations in $R(E_T)$ at
large $E_T$ of a
comparable size. These fluctuations, which are generally regarded as spurious,
are an example of
systematic errors that do not cancel when taking the ratio of $J/\psi$ and $DY$
cross-sections and
have to be understood. Third, the failure of the model to describe
quantitatively the ratio
$R(E_T)$ in the region $E_T \ \lsim \ 100$ GeV is in sharp contrast with the
conclusions reached
in Section 3 from a direct comparison of the model results with the $E_T$
distribution of the
$J/\psi$ which showed reasonable agreement in this $E_T$ region. In order to
understand the origin of
this contradiction we have plotted in Fig. 8 the theoretical curve of Fig. 7
(full curve), and
compared it with the $J/\psi$ suppression obtained from the ratio
$\overline{R}(E_T)$ of the
experimental $E_T$ distribution of the $J/\psi$ (for the 1996 NA50 data) over
the theoretical one
for the $DY$ (full curve of Fig. 5). We see that the agreement in shape between
theory and
experiment has considerably improved; the remaining drop, strongly
reduced compared with that of $R(E_T)$, can be
attributed to a local minimum in the $J/\psi$ cross-section at $E_T\sim
55$ GeV (difficult to explain in any model,
see comments at the end of the previous
Section).
The ratio $\overline{R}(E_T)$ is now
rather well
described in the region $E_T < 100$ GeV -- except for the first $E_T$ bin.
Moreover, in the ratio
$\overline{R}(E_T)$ the break at $E_T \sim 55$ GeV has practically disappeared.
 Fig. 8 indicates that the ratio $R(E_T)$ is very sensitive to the shape of the
$DY$ distribution.
% CAMBIO
%,
%and,  together with the results of Section 3, it suggests that the break in
%$R(E_T)$ is
%mostly due to fluctuations in the $DY$ distribution (and partly to the local
%minimum found at
%$\sim 55$ GeV in the $J/\psi$ distribution), rather than to a genuine break in
%the $J/\psi$ one.
We would like to stress that Fig. 8 contains no new information. However, 
it is useful since its comparison with Fig. 7 shows the effect on the ratio 
$R(E_T)$ of smoothing out the $DY$ cross-section.

Before concluding this section we would like to comment on the modifications in
the ratio $R(E_T)$
when using the comover density computed in the WNM, rather than the one based
on DPM, Eq.
(\ref{8e}). As discussed in Section 2, the WNM underestimates the number of
negative particles in
a central $PbPb$ collision at $y^* \sim 0$ by $15\div 30$ \%. The corresponding
DPM value is 30 \%
larger than the WNM one and in better agreement with the NA49 data (see Section
2). If we would
decrease the density of comovers by 30 \% for the most central $E_T$ bin, the
value of $R(E_T)$
would decrease by about 10 \%. Actually, the net effect would be significantly
smaller, since the
WNM multiplicity is smaller than the DPM one also in $SU$, and this can be
compensated by a
corresponding increase of $\sigma_{co}$. Although a difference would remain, it
would not
basically change the conclusions of the present analysis. On the contrary, our
results would be
changed if we were to use the WNM in the calorimeter rapidity region, in order
to determine the
$E_T - E_{ZDC}$ correlation. In this case we would obtain an $E_T$ distribution
for the $J/\psi$
which would be too large in the upper half of the $E_T$ interval.

\section{Transverse momentum broadening}
\hspace*{\parindent}
A mechanism producing an increase of the $\langle p_T\rangle $ of any type of
particle produced in
$pA$ or $AB$ collisions with increasing nuclear sizes or centrality has been
known for a long time
\cite{26r,27r}. It is due to initial state rescattering. More precisely, for
$J/\psi$ production
this increase is due to rescattering of the projectile and target gluons,
before fusion, with
target and projectile nucleons respectively, encountered in their path through
the nuclei. The
average broadening of the intrinsic gluon distribution in each collision is
denoted by $\delta_0$.
In Ref. \cite{27r} it has been shown that the $p_T$ broadening of the $J/\psi$
is affected by
$J/\psi$ absorption. In particular the suppression in $PbPb$ collisions in a
deconfining approach
\cite{4r}, produces a maximum in the $E_T$ dependence of $\langle p_T^2\rangle
_{J/\psi}$ at $E_T
\sim 100$ GeV followed by a decrease with increasing $E_T$. This peculiar
behavior has been
considered as a signature of Quark-Gluon Plasma formation.

In this section, we follow the formalism of $p_T$ broadening of the $J/\psi$ in
Ref. \cite{27r},
but using absorption by comovers instead of the one due to deconfinement.

The broadening of $p_T$ is given by \beq \label{15e} \delta_{AB}(b) \equiv
\langle p_T^2\rangle
_{AB}(b) - \langle p_T^2\rangle _{pp} = N_{AB}(b) \ \delta_0\ \ . \eeq $N_{AB}$
is the average
number of collisions of the projectile and target gluons with target and
projectile nucleons
respectively, up to the formation point of the $c\overline{c}$ pair, at fixed
$b$. This point is
specified by the impact parameter ${b}$ and the positions $({s}, z)$ and $({b}
- {s} , z')$ in the
two nuclei. One has \beq \label{16e} N_{AB}({b}, {s}, z, z') = \sigma_{gN}A
\int_{- \infty}^{z}
dz_A \  \rho_A ({s}, z_A) + \sigma_{gN}B \int_{- \infty}^{z'} dz_B \ \rho_B
({b} - {s}, z_B) \ \ .
\eeq Here $\sigma_{gN}$ is the gluon-nucleon cross section. This expression
(\ref{16e}) has to be
averaged over all positions of the $c\overline{c}$ formation point with a
weight given by the
product of nuclear densities and survival probabilities: \beq \label{17e}
W({b}, {s}, z, z') =
\rho_A ({s}, z) \ \rho_B({b}-s, z') S_A ({s}, z) \  S_B ({b} - {s}, z') \
S^{co}({b}, {s})\ \  ,
\eeq where \beq \label{18e} S_A({b}, z) = \exp \left ( - A \sigma_{abs}
\int_z^{\infty} d\tilde{z}
\ \rho_A(b, \tilde{z}) \right ) \eeq is the survival probability due to nuclear
absorption
\cite{13r} and $S^{co}(b, s)$ is the survival probability due to interaction
with comovers, Eq.
(\ref{2e}). The latter does not depend on the $c\overline{c}$ formation point.

We obtain in this way \beq \label{19e} N_{AB}(b) = {\int d^2 {s} \int_{-
\infty}^{+ \infty} dz
\int_{- \infty}^{+ \infty} dz' \ W({b}, {s}, z, z') \ N_{AB}({b}, {s}, z, z')
\over \int d^2{s}
\int_{- \infty}^{+ \infty} dz \int_{- \infty}^{+ \infty} dz' \  W({b}, {s}, z,
z')} \ \ . \eeq
This expression can be written after some transformations as \beq \label{20e}
N_{AB}(b) =
\sigma_{gN}\  {\int d^2{s} \ S^{co}({b}, {s}) \left [ N_A ({s}) D_B({b} - {s})
+ N_B ({b} - {s})
D_A({s}) \right ] \over \int d^2{s} \ S^{co}({b}, {s}) \ D_A({s}) \ D_B({b} -
{s})}\ \ , \eeq where
\beq \label{21e} D_A({s}) = {1 \over A \sigma_{abs}} \left ( 1 - \exp \left [-
A \sigma_{abs} \
T_A({s}) \right ] \right ) \eeq and \beq \label{22e} N_A({s}) = {1 \over A
\sigma_{abs}^2} \left (
\sigma_{abs} A T_A(s) - 1 + \exp \left [- A \sigma_{abs} T_A({s}) \right ]
\right ) \ \ , \eeq
$D_B=D_A(A\rightarrow B)$ and $N_B=N_A(A\rightarrow B)$.

Finally, the corresponding quantity at fixed transverse energy can be obtained
as \beq \label{23e}
N_{AB} (E_T) = {\int d^2b \ P(E_T, b) \ \sigma_{AB}(b) \ N_{AB}(b) \over \int
d^2b \ P(E_T, b) \
\sigma_{AB}(b)}\ \ , \eeq where $P(E_T, b)$ is the $E_T-b$ correlation
function, Eq. (\ref{10e}),
and $\sigma_{AB}(b)$ is given by Eq. (\ref{5e}).

The values of $\langle p_T^2\rangle _{pp}$ and $\sigma_{gN} \delta_0$ at 158
AGeV/c are obtained
from a fit to the NA50 data \cite{10r}. One obtains $\langle p_T^2\rangle _{pp}
= 1.03 \div 1.10$
(GeV/c)$^2$ and $\sigma_{gN} \delta_0 = 0.39 \div 0.47\simeq 10.0\div 12.1$
(GeV fm)$^2$
(depending on whether the effect of comovers is included or not). The value of
$\sigma_{gN}
\delta_0=0.39$ we obtain for nuclear absorption with $\sigma_{abs}=7.3$ mb and
no comovers, agrees
with that obtained in \cite{27r}, $9.4\pm 0.7$ (GeV fm)$^2$, from a fit to $pA$
and $SU$ data
\cite{28br,28bbr}. As suggested in \cite{gh}, we   should take different values
of $\langle
p_T^2\rangle _{pp}$ in $SU$ and $PbPb$, since this value increases with energy.
Using the values
measured \cite{28r} in $\pi^- p$ collisions at 150 and 200 GeV/c, the value
$\langle
p_T^2\rangle_{pp} = 1.07$ (GeV/c)$^2$ would correspond to $\langle
p_T^2\rangle_{pp}  = 1.23$
(GeV/c)$^2$ at 200 GeV/c. This last value coincides with the one measured in
Ref. \cite{28r},
$1.23 \pm 0.05$ (GeV/c)$^2$, in $pp$ collisions at 200 GeV/c. (From a fit to
$DY$ data in $pA$,
$OCu$, $OU$ and $SU$ \cite{28br}, a value of $\sigma_{gN} \delta_0=0.13$ is
obtained, whose ratio
over the values for $J/\psi$ is $\sim 0.33$, smaller than the value $4/9\simeq
0.44$ suggested
\cite{26r} by the difference of coupling between gluons and quarks or gluons.)

Our results for nuclear absorption plus comovers with $\sigma_{abs}=6.7$ mb and
$\sigma_{co}=0.6$
mb (nuclear absorption alone with $\sigma_{abs}=7.3$ mb), for $SU$ with
$\langle
p_T^2\rangle_{pp}  = 1.23$ (GeV/c)$^2$ and $\sigma_{gN} \delta_0 = 0.42$
(0.40), and for $PbPb$
collisions with $\langle p_T^2\rangle_{pp}  = 1.10$ (GeV/c)$^2$ (in agreement
with the mentioned
rescaling between 200 and 158 AGeV/c) and the same  $\sigma_{gN} \delta_0$ as
in $SU$, are shown
in Fig. 9 and compared with experimental data \cite{10r,28pbr}. We see that, in
$PbPb$ collisions with
comovers, there is a small maximum at $E_T \sim 125$ GeV. However, after this
maximum, $\langle
p_T^2\rangle_{AB}$ is practically constant and only slightly smaller than the
one obtained with
nuclear absorption alone. This is in contrast with the sharper decrease at
large $E_T$ found in a
deconfining scenario \cite{27r}. 
%%% CAMBIO
The physical origin of this decrease is the same in both approaches. At large 
$E_T$, corresponding to large comover or energy density,
the $J/\psi$, with large $\langle
p^2_T\rangle$ due to a large number of initial $gN$ collisions,
is suppressed by either the comover or the deconfining 
mechanisms. However, this effect turns out to be numerically much lower in the
former approach.
%%%%%%
Unfortunately, with the present data it is not
possible to
clearly discriminate between these two predictions.

\section{Conclusions and prospects}
\hspace*{\parindent}
We have presented a direct comparison of the available NA50 data for the $E_T$
distribution of the
$J/\psi$ with the results obtained in a conventional framework based on nuclear
absorption plus
comover interaction. 
%The agreement of the model results with the data is
%reasonably good in
%the region $E_T < 100$ GeV. In this region the data show small
%fluctuations
%around the theoretical curve but there is no sign of a break at any value of
%$E_T$.
%The comparison of the model results with the
%data for the ratio $R(E_T)$ of $J/\psi$ over $DY$
%cross-sections is not so good. In particular, the data seem to have a break at
%$E_T \sim 55$ GeV which is
%not present in the model. However, this break disappears when plotting the
%ratio of the experimental $E_T$
%distribution of the $J/\psi$ divided by the theoretical one of the $DY$. \par

Our analysis suggests that 
%this break is due to fluctuations in the $DY$
%distribution rather than
%to a break in the $J/\psi$ one and that 
the presence of new physics in the
region $E_T < 100$ GeV
is unlikely. On the contrary, the region $E_T > 100$ GeV is very interesting
and should be studied
with great care in the 1998 high statistics run. Agreement of the $J/\psi$
cross-section with the
comovers results for $E_T \ \lsim \ 100$ GeV together with a
si\-gni\-fi\-can\-tly sharper decrease
for $E_T > 100$ GeV (for which there might be some hint in the 1996 data), would
signal the onset of a
truly anomalous $J/\psi$ suppression.

Is it possible with the present data to distinguish a deconfining phase
transition scenario from
the more conventional one described here? In order to answer this question we
have to distinguish
between deconfining scenarios producing sharp breaks in the ratio $R(E_T)$
\cite{11r,5r} from
others leading to a smooth behavior of this ratio \cite{2r,3r,4r,5r}. For the
former, a clear-cut answer
will probably come from the 1998 data. On the contrary, it will be more
difficult to
distinguish the second type of deconfining models from the comover approach
presented here.

A very clear way to do so would be to show that the onset of the anomalous
suppression is abrupt,
i.e. it is not present below some critical density -- for instance the maximal
one reached in $SU$
collisions \cite{2r}. Up to a recent date, there was some evidence for that
\cite{10r}. Indeed, the
effect of the comovers in $SU$ produced a somewhat larger suppression
\cite{7r,8r} than the measured
one. At present, however, the experimental errors in the ratio $R(E_T)$ in $SU$
collisions have
been increased by a factor 2.8 (see Ref. \cite{30r}; also the experimental
errors for $PbPb$ have
increased, by a factor 1.4, which has been taken into account in this work). In
view of
that, it is no longer possible to claim that the $J/\psi$ suppression in $SU$
is too large in the
comover approach \cite{8r} or that the onset of the anomalous suppression is an
abrupt
one.

As we have shown in Section 5, there is a difference between comovers and
deconfining scenarios
regarding the behavior of $\langle p_T^2\rangle $ versus $E_T$. According to
Ref. \cite{27r}, in a
deconfining scenario this quantity has a maximum at $E_T \sim 100$ GeV and
decreases at larger
$E_T$ values. In the comover approach presented here, this drop is practically
absent and the $E_T$
dependence is close to the one obtained with nuclear absorption alone. Although
no such drop is
seen in the data, the present experimental errors are rather large and a clear
conclusion is not
possible.

A promising possibility is the measurement of the $J/\psi$ suppression at
higher energies.
The $J/\psi$ suppression due to either comover interactions or deconfinement,
is expected to
increase substantially with increasing energy. In the first case, this is due
to the increase of
the density of comovers with increasing energy. In the second case, it is due
to the corresponding
increase of energy density -- while the critical value of this quantity is
unchanged. Therefore,
it is important to make predictions at higher energies in both approaches,
using the values of the
parameters determined from present data. One can hope that the differences in
the predictions of
the two approaches will be sufficiently large to be experimentally measurable.

The main uncertainty in the determination of the absolute value of the
suppression at high
energies resides in the value of $dN/dy$ at $y^* \sim 0$. For instance, in
central $PbPb$
collisions, at RHIC energies, one expects in DPM a value \cite{31r} for
negative particles
$dN^-/dy|_{y^*\sim 0} = 1000$, and 3500 at $\sqrt{s_{NN}} = 5.5$ TeV. On the
contrary, from the
scaling in the number of participants (WNM) one expects a value $dN^-/dy|_{y^*
\sim 0} = 400$ at
$\sqrt{s_{NN }} = 200$ GeV and 800 at $\sqrt{s_{NN}} = 5.5$ TeV. In the first
case, there is an
increase by roughly a factor 5 at RHIC (17.5 at LHC) with respect to the value
at $\sqrt{s_{NN}} =
17$ GeV. In the second case, there is only an increase by a factor 2 at RHIC (4
at LHC), which is
due to the corresponding increase of $dN/dy|_{y^* \sim 0}$ in $pp$ collisions.
An estimate at
RHIC (LHC) of the $J/\psi$ survival probability in central $PbPb$ collisions is
given in Table 1.
The numbers in this Table, for comover absorption alone, are obtained from Eq.
(\ref{2e}) by
rising the comover absorption, computed at $\sqrt{s_{NN}}=17$ GeV for a central
$E_T$
bin ($E_T\sim 145$ GeV), to a power 2 (4) in the case of the WNM and to a power
5 (17.5) in the
case of DPM. The corresponding numbers for the total $J/\psi$ suppression are
obtained by multiplying the ones
for comovers alone given in Table 1, by the nuclear absorption. The latter is
expected to depend little on energy
\cite{32r}.

These estimates illustrate the important increase of the $J/\psi$ suppression
with e\-ner\-gy and
also the dramatic uncertainties associated to the value of $dN/dy|_{y^* \sim
0}$. Clearly, a more
detailed calculation is needed which takes into account the modifications of
parton densities
inside nuclei (usually neglected at SPS energies) and also the changes in the
Glauber formulae due
to the increase with energy of $\sigma_{pp}$. However, it is obvious that the
$J/\psi$
suppression will increase strongly with increasing energies and it is very
unlikely that the
results will be the same in the comover and in the deconfining frameworks. \par
\vskip 5 truemm

\noindent {\bf Acknowledgments:}
It is a pleasure to thank A. B. Kaidalov, A. Krzywicki, C. A. Salgado, Yu. M.
Shabelski and J.
Tr$\hat{{\mbox a}}$n Thanh V$\hat{{\mbox a}}$n for discussions, and F.
Bellaiche, B. Chaurand, C.
Gerschel, M. Gonin, C. Louren\c co and A. Romana for discussions and useful
information on the
experimental data. N. A. thanks Direcci\'on General de Investigaci\'on
Cient\'{\i}fica y T\'ecnica
of Spain for financial support and J. Hern\'andez, M. Mart\'{\i}nez, J. Puga,
J. Rold\'an and J.
Terr\'on for discussions on experimental aspects. E. G. F. thanks Fundaci\'on
Ram\'on Areces of
Spain for financial support. A. C. acknowledges partial support from NATO grant
OUTR.LG971390 and
INTAS grant 93-79. Laboratoire de Physique Th\'eorique et Hautes Energies is
Laboratoire associ\'e
au Centre National de la Recherche Scientifique -- URA D00063.

\newpage
\centerline{\bf Table captions:}
\vspace{1cm}

\noi {\bf Table 1.}
Comover
and total $J/\psi$ suppression at SPS, RHIC and LHC, for central $PbPb$
collisions, in the WNM and the DPM (see text for an explanation).

\vspace{7cm}
\centerline{\bf Table 1}
\vspace{1cm}
\begin{center}
\begin{tabular}{ccccccc}
\hline \hline
 & SPS & SPS & RHIC & RHIC & LHC & LHC \\
 & Comover & Total & Comover & Total & Comover & Total\\ \hline
WNM & 0.62 & 0.23 & 0.38 & 0.14 & 0.14 & 0.06 \\
DPM & 0.62 & 0.23 & 0.09 & 0.03 & $2\cdot 10^{-4}$ & $8\cdot
10^{-5}$\\ \hline \hline \\
\end{tabular}
\end{center}

\newpage
\centerline{\bf Figure captions:}
\vspace{1cm}

\noi {\bf Figure 1.} Rapidity distribution of negative hadrons in central
$PbPb$ collisions at 158
AGeV/c. Preliminary data of the NA49 Collaboration \cite{20r} (black circles)
are compared with
the DPM results (solid line) using the same centrality criterium, and with the
scaling in the
number of participants (dashed line \cite{20r}). \\

\noi {\bf Figure 2.} $E_T - E_{ZDC}$ correlation: the full line is obtained in
DPM from Eqs.
($9^{\prime \prime}$) and
(\ref{9e}) and gives a very good description of the NA50 collaboration
data \cite{10r}.
The dotted line is obtained in DPM from Eqs. (\ref{8e}) and (\ref{9e}). The
dashed line is obtained
in the WNM with $E_T(b) = 0.4$ $[m_A(b) + m_B(b)]$ GeV \cite{4r}.\\

\noi {\bf Figure 3.} Inclusive cross-section $d\sigma^{DY}/dE_T$ for $DY$ pair
production with
$M_{\mu \mu}> 4.2$ GeV/c$^2$ from the 1995 NA50 data \cite{24r} compared to the
results obtained
from Eq. (\ref{11e}) with $\sigma_{abs} = \sigma_{co} = 0$. The full curve is
obtained in DPM and the dashed one in the WNM. The
normalization constant
$\sigma_{pp}^{DY}/\sigma_{pp}$  in Eq. (\ref{3e}) is $9\cdot 10^{-10}$. Note
that, in order to
compare with the experimental value of $\sigma_{pp}^{DY}/\sigma_{pp}$, this
normalization factor
should be divided by 5 due to the $E_T$ binning in Fig. 3.  \\

\noi {\bf Figure 4.} Preliminary $E_T$ distribution $dN^{DY}/dE_T$ for $DY$
pair production with
$M_{\mu \mu}> 4.2$ GeV/c$^2$ for the 1996 NA50 data \cite{10r} compared to the
theoretical curves
of Fig. 3 normalized to the data. The common normalization factor
is 0.10
fm$^{-2}$. \\

\noi {\bf Figure 5.} Inclusive cross-section $d\sigma^{J/\psi}/dE_T$ for
$J/\psi$ production from
the 1995 NA50 data \cite{24r} compared with the results obtained from Eq.
(\ref{11e}). The
normalization constant $B_{\mu \mu}\sigma_{pp}^{J/\psi}/\sigma_{pp}$ in Eq.
(\ref{3e}) is
$2.4\cdot 10^{-7}$. The dotted line is obtained with nuclear absorption alone
($\sigma_{abs} =
7.3$ mb, $\sigma_{co} = 0$), while the solid line contains the effect of
comovers with
$\sigma_{abs}= 6.7$ mb and $\sigma_{co} = 0.6$ mb. The dashed line is obtained
in the WNM with
nuclear absorption alone ($\sigma_{abs} = 7.3$ mb, $\sigma_{co} = 0$). \\

\noi {\bf Figure 6.} Premiminary $E_T$ distribution $dN^{J/\psi}/dE_T$ for
$J/\psi$ production from
the 1996 NA50 data \cite{10r}, compared with the theoretical curves of Fig. 5
normalized to the
data. The common normalization factor is 5.57 fm$^{-2}$.
% CAMBIO 4
The circles and the crosses 
correspond to
two
different experimental methods \cite{10r}: fitting
and counting procedures. \\

\noi {\bf Figure 7.} The ratio $R(E_T)$ of $J/\psi$ over $DY$ versus $E_T$ both
from the 1995
\cite{1r} (open symbols) and preliminary 1996 \cite{10r} (black symbols) NA50
data compared to the
ratio of theoretical curves (solid lines) in Figs. 4 and 6 (with comovers,
solid line).  The
dotted line is obtained in DPM with nuclear absorption alone ($\sigma_{abs} =
7.3$ mb, $\sigma_{co}
= 0$). The normalization factor
(61.2) is the one obtained in \cite{8r} from a fit to
the $pA$, $SU$ and
$PbPb$ data. This normalization coincides with the one obtained from the
normalizations of the
individual $E_T$ distributions in Figs. 4 and 6 after correcting the latter for
the different
$E_T$ binnings, the experimental acceptances and the different $DY$ mass range
-- which is $2.9 <
M < 4.5$ GeV/c$^2$ in the ratio $R(E_T)$ and $M > 4.2$ GeV/c$^2$ in Fig. 4.\\

\noi {\bf Figure 8.} The theoretical curve of Fig. 7 (solid line) is compared
to the ratio
$\overline{R}(E_T)$ of the experimental $E_T$ distribution of Fig. 6 over the
theoretical $DY$
distribution of Fig. 4 (solid line). Here the normalization of
$\overline{R}(E_T)$ is arbitrary
-- since we are only interested in the change in the shape of $R(E_T)$ when
smoothing the $DY$
$E_T$ distribution. \\

\noi {\bf Figure 9.} $\langle p_T^2\rangle_{AB}$ in a) $SU$
and b) $PbPb$ collisions at SPS. Solid line: nuclear
absorption plus comovers; dotted line: nuclear absorption alone (see
text for the values of the corresponding parameters). Black circles are
experimental data from \cite{10r,28pbr}.

\newpage

\centerline{\bf Figure 1}
\vspace{1cm}

\hspace{-1.2cm}\epsfig{file=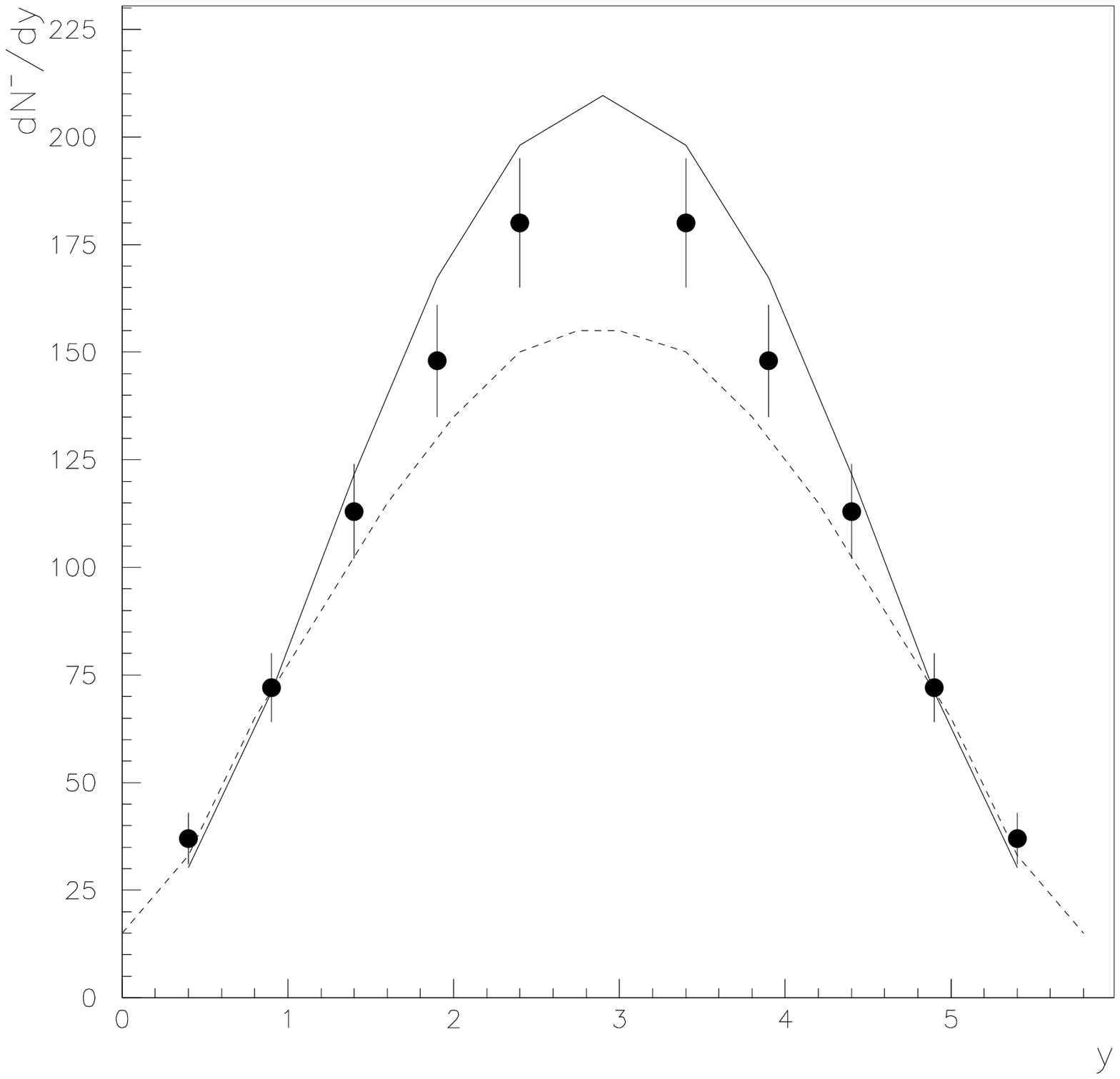,width=16.cm}

\newpage

\centerline{\bf Figure 2}
\vspace{1cm}

\hspace{-1.2cm}\epsfig{file=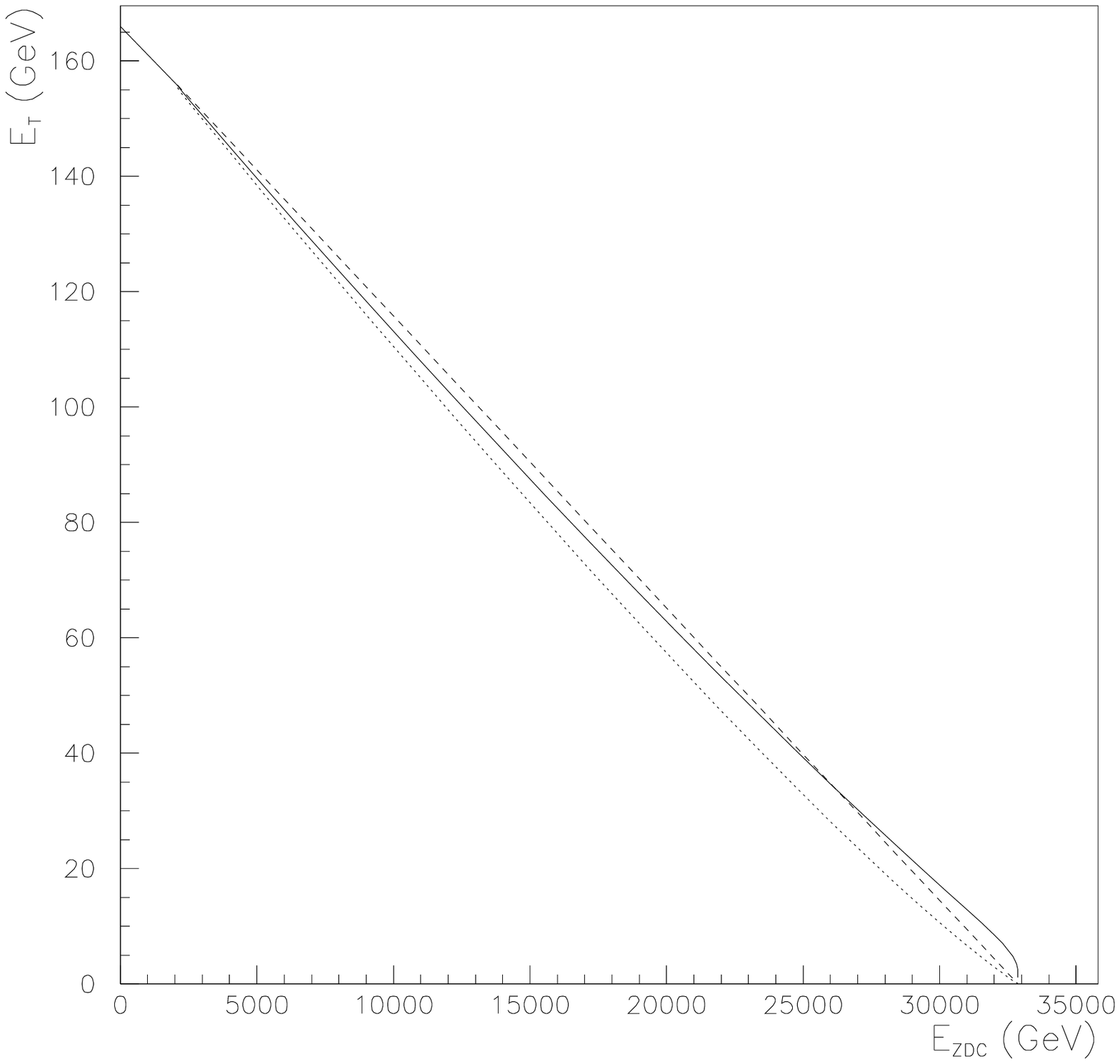,width=16.cm}

\newpage

\centerline{\bf Figure 3}
\vspace{1cm}

\hspace{-1.2cm}\epsfig{file=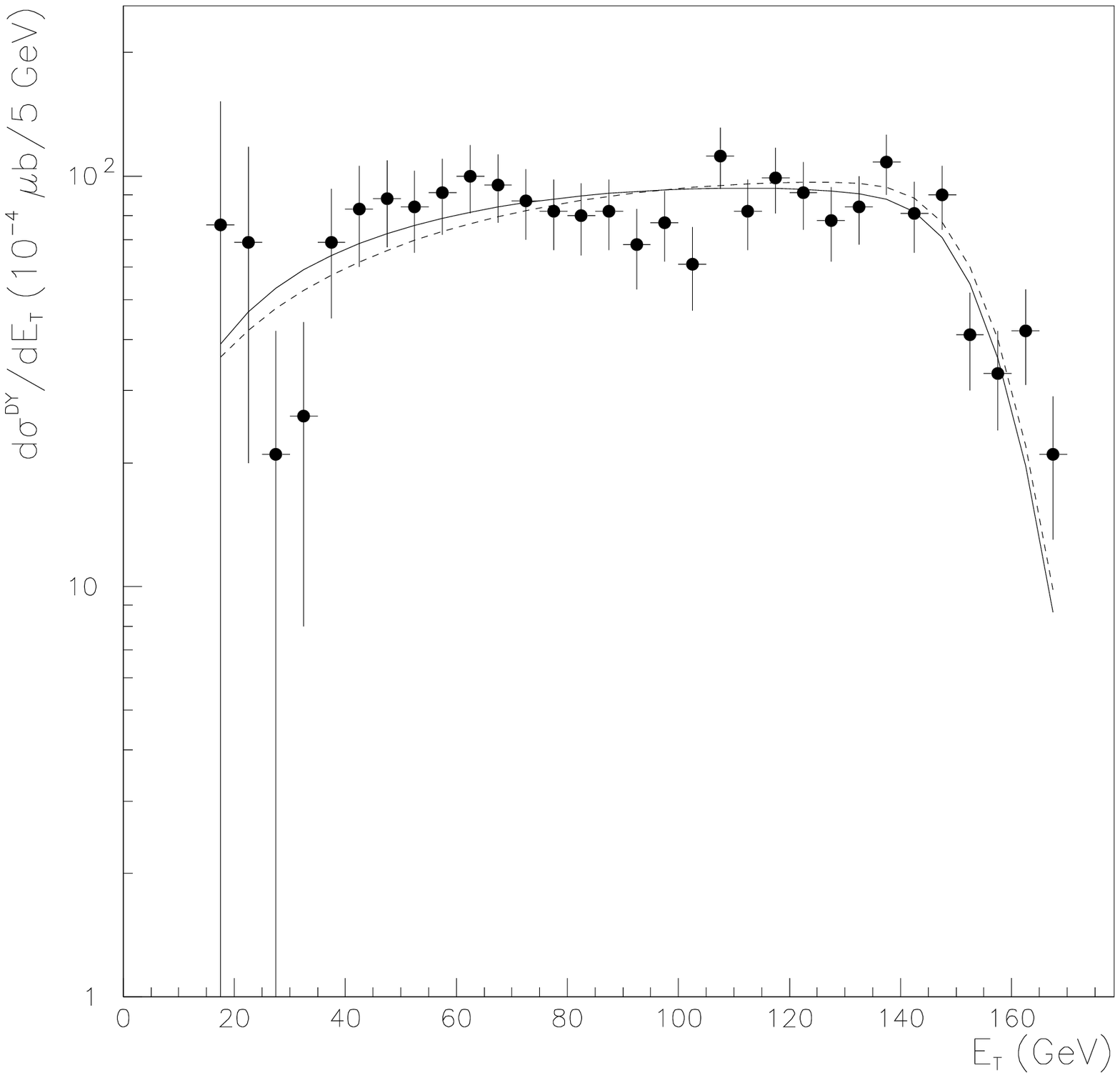,width=16.cm}

\newpage

\centerline{\bf Figure 4}
\vspace{1cm}

\hspace{-1.2cm}\epsfig{file=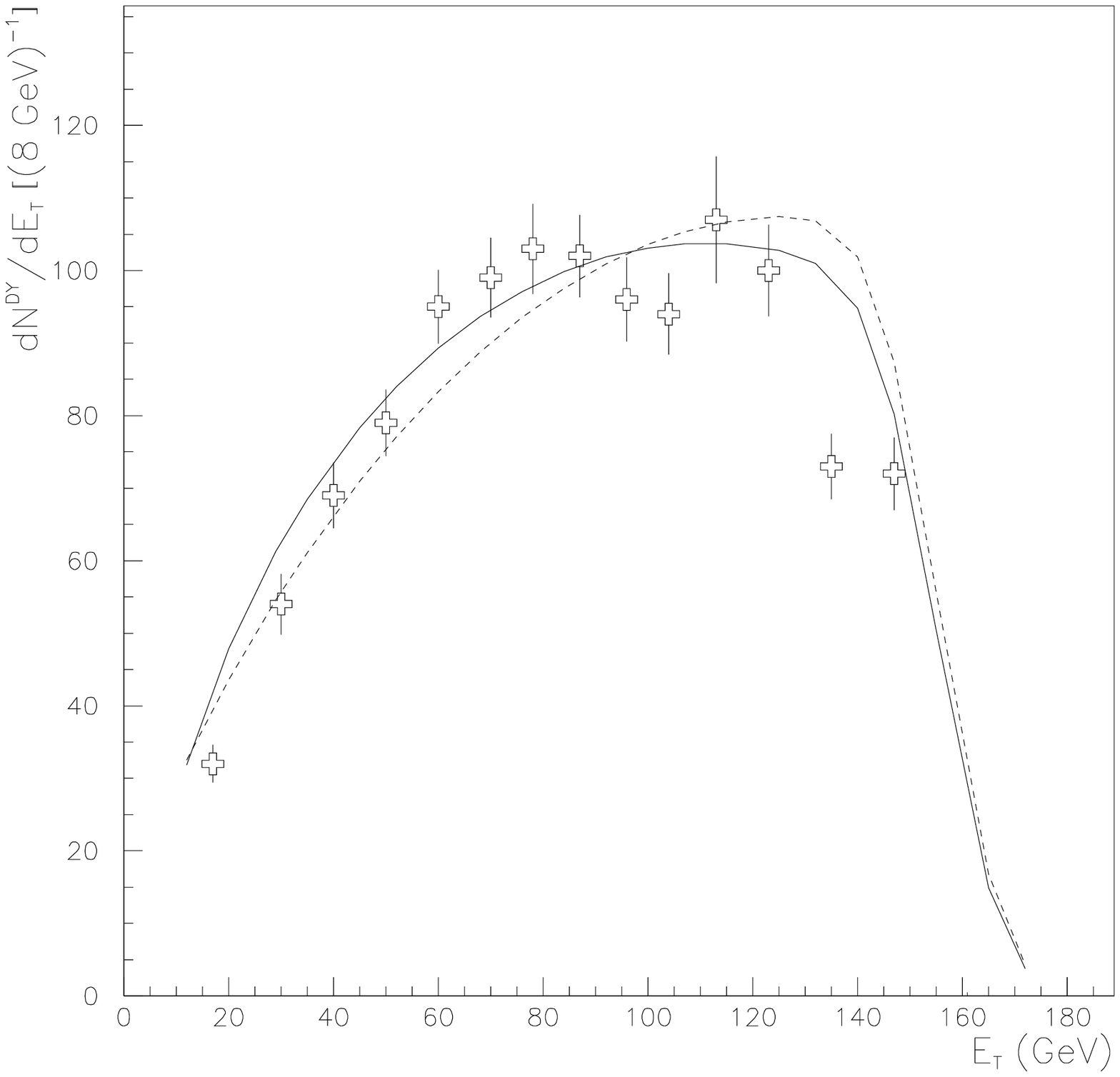,width=16.cm}

\newpage

\centerline{\bf Figure 5}
\vspace{1cm}

\hspace{-1.2cm}\epsfig{file=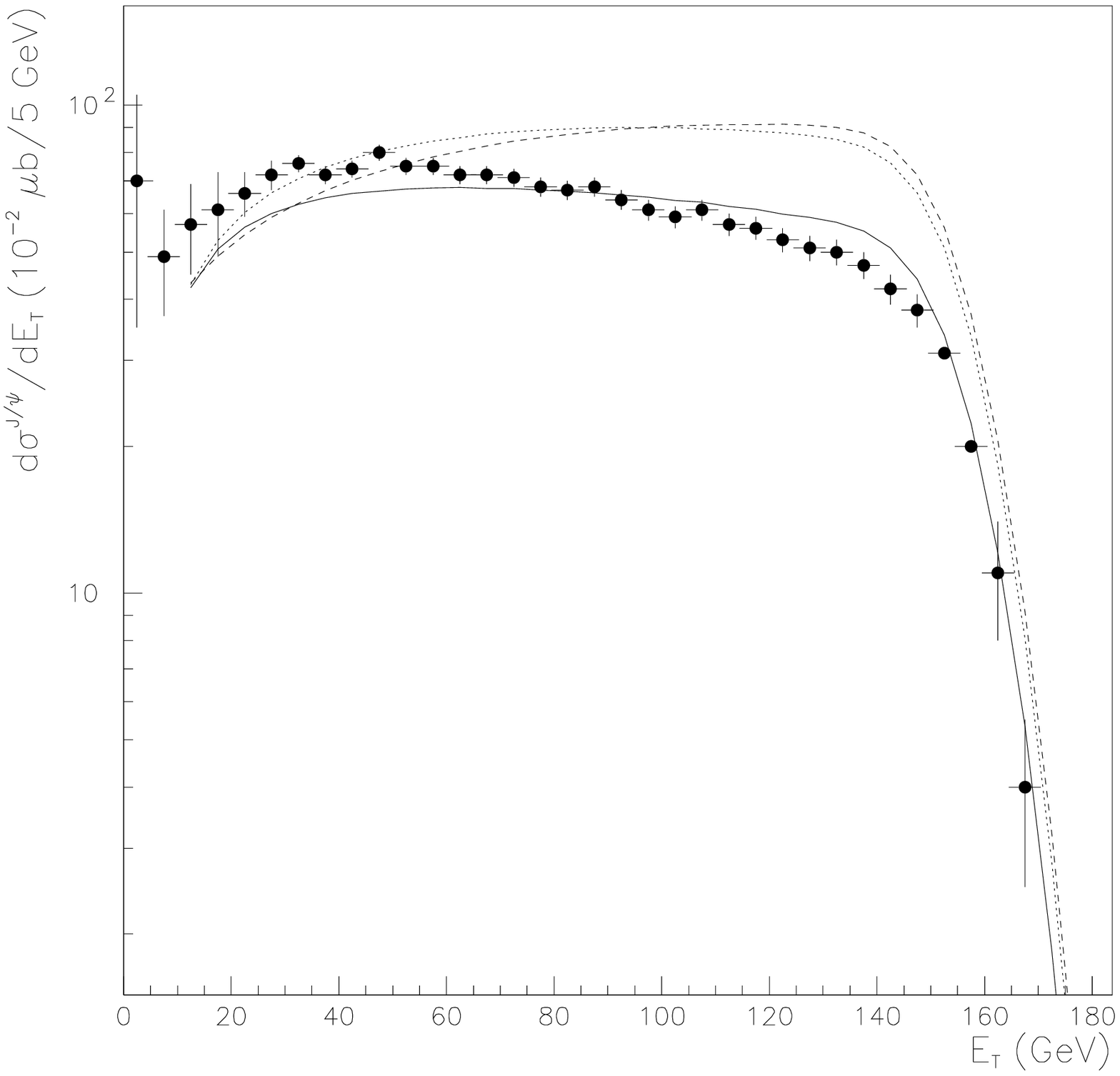,width=16.cm}

\newpage

\centerline{\bf Figure 6}
\vspace{1cm}

\hspace{-1.2cm}\epsfig{file=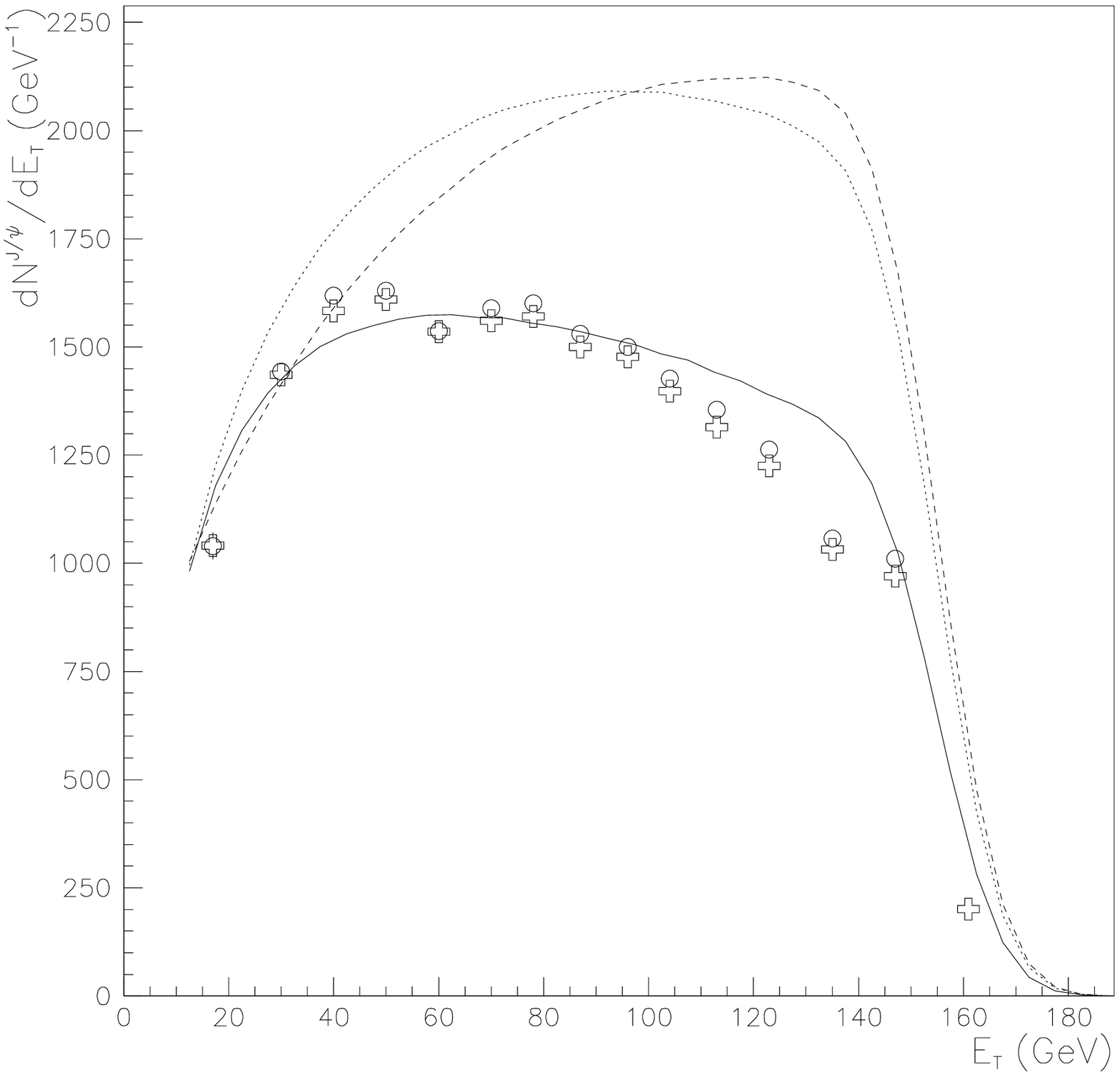,width=16.cm}

\newpage

\centerline{\bf Figure 7}
\vspace{1cm}

\hspace{-1.2cm}\epsfig{file=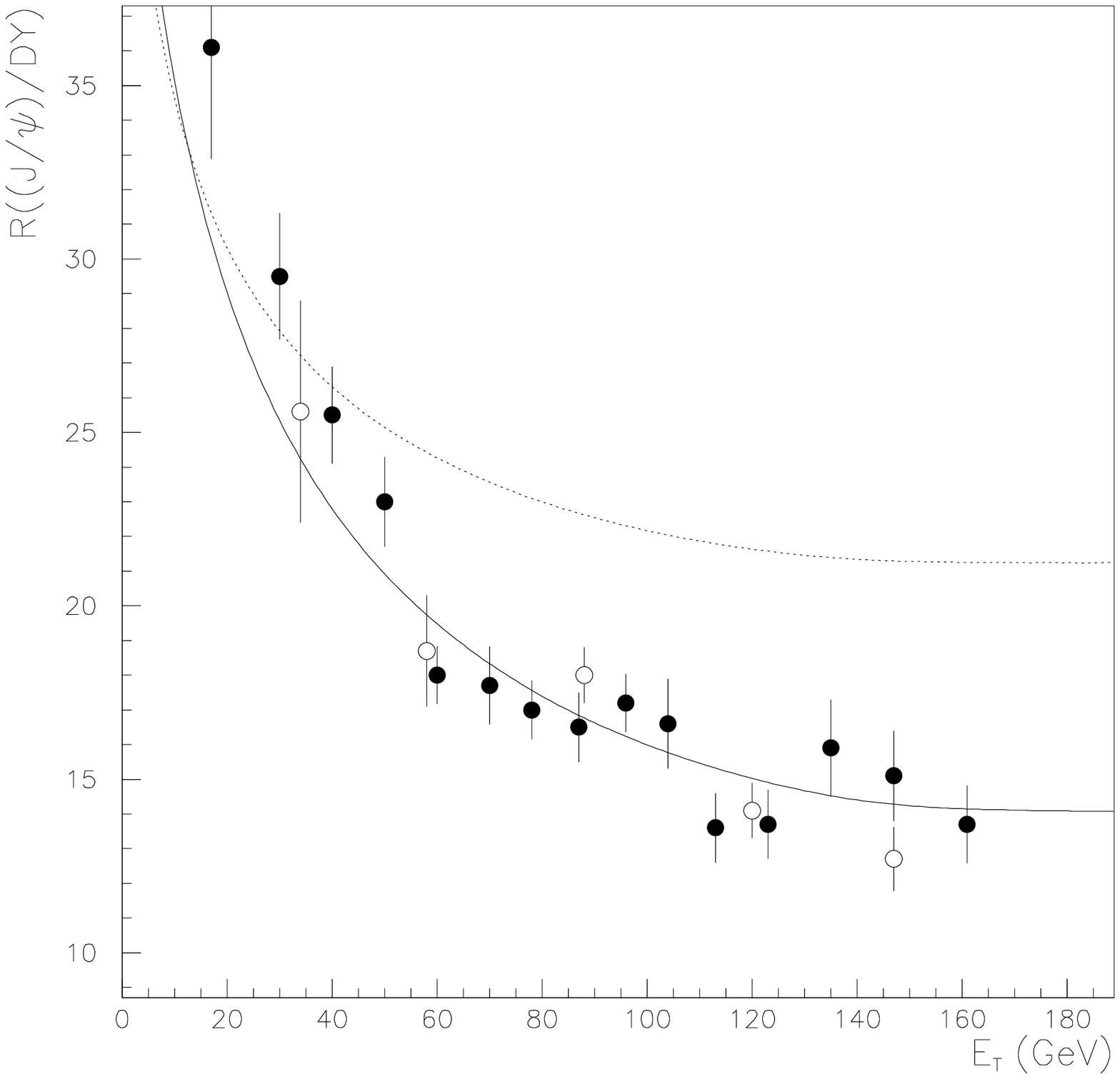,width=16.cm}

\newpage

\centerline{\bf Figure 8}
\vspace{1cm}

\hspace{-1.2cm}\epsfig{file=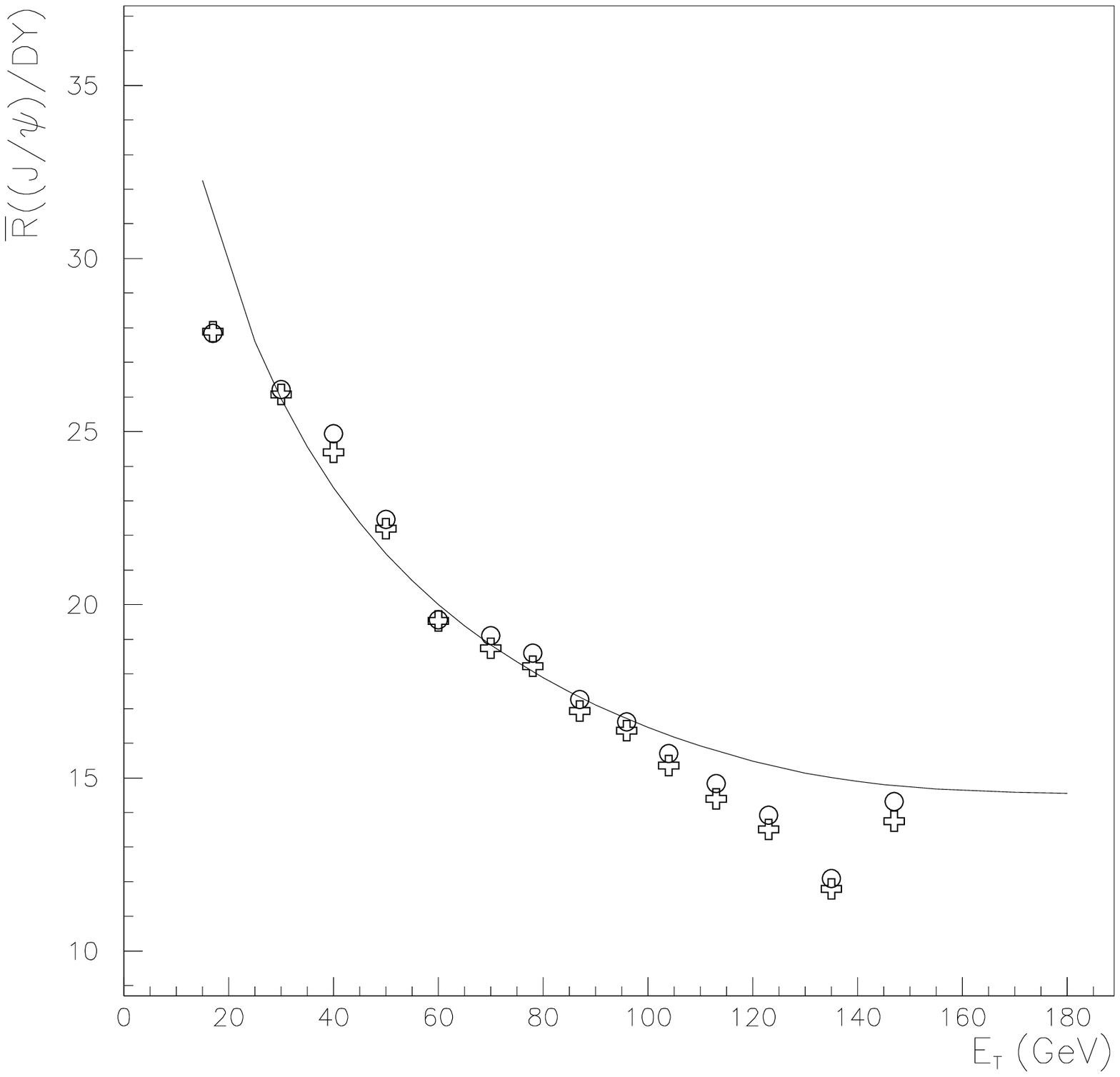,width=16.cm}

\newpage

\centerline{\bf Figure 9}
\vspace{1cm}

\hspace{-1.2cm}\epsfig{file=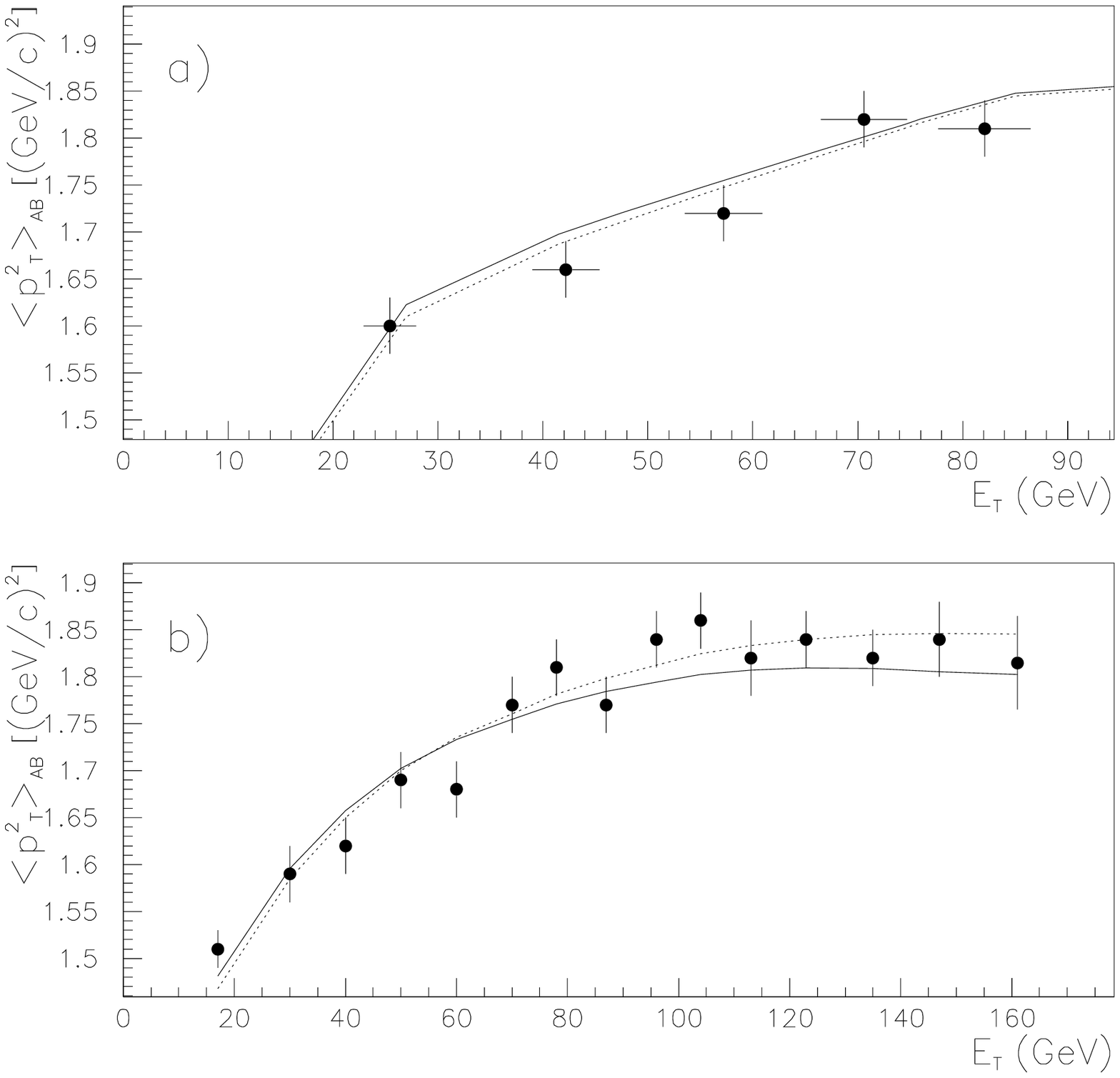,width=16.cm}


\newpage

\begin{thebibliography}{99}

\bibitem{1r} NA50 Collaboration: M. C. Abreu {\it et al.}, Phys. Lett. {\bf
B410}, 327 (1997); 337.

\bibitem{1br} T. Matsui and H. Satz, Phys. Lett. {\bf B178}, 416 (1986).

\bibitem{2r} J.-P. Blaizot and J.-Y. Ollitrault, Phys. Rev. Lett. {\bf 77},
1703 (1996); Nucl.
Phys. {\bf A610}, 452c (1996).

\bibitem{3r} C.-Y. Wong, Nucl. Phys. {\bf A610}, 434c (1996); Phys. Rev. {\bf
C55}, 2621 (1997).

\bibitem{4r} D. Kharzeev, C. Louren\c co, M. Nardi and H. Satz,
Z. Phys. {\bf C74}, 307 (1997).

\bibitem{5r} R. Vogt, Phys. Lett. {\bf B430}, 15 (1998).
In this
article it is claimed that nuclear absorption plus comover interaction cannot
explain the size of
the $J/\psi$ suppression observed in $PbPb$ collisions. An opposite conclusion
is reached in
\cite{8r} and in the present work. In [6] the Wounded Nucleon Model is used to
compute the
comovers suppression as well as the $E_T - b$ correlation.  Another difference
resides in the
value of the absorptive cross-section -- which is taken to be equal to 4.8 mb
in Ref. \cite{5r}
and 6.7 mb in our case.

\bibitem{6r} S. Gavin and R. Vogt, Nucl. Phys. {\bf A610}, 442c (1996); Phys.
Rev. Lett. {\bf 78},
1006 (1997). One of the authors (R.V.) has claimed \protect{\cite{5r}} that
this work has an error
in the relative normalizations in $SU$ and $PbPb$ collisions which invalidates
its conclusions.

\bibitem{7r} A. Capella, A. B. Kaidalov, A. Kouider Akil and C.
Gerschel, Phys. Lett. {\bf B393}, 431 (1997).

\bibitem{8r} N. Armesto and A. Capella,
Phys. Lett. {\bf B430}, 23 (1998); J. Phys. {\bf G23}, 1969
(1997); N. Armesto, in
Proceedings of the XXXIInd Rencontres de Moriond, edited by J. Tr$\hat{{\mbox
a}}$n Thanh
V$\hat{{\mbox a}}$n (Editions Fronti\`eres, Paris, 1997), p. 519.

\bibitem{9r} W. Cassing and C. M. Ko, Phys. Lett. {\bf B396}, 39 (1997); W.
Cassing and E. L.
Bratkovskaya, in Proceedings of the XXXIInd Rencontres de Moriond, {\it ibid},
p. 531; Nucl. Phys.
{\bf A623}, 570 (1997); J. Geiss, C. Greiner, E. L. Bratkovskaya, W.
Cassing and U. Mosel, nucl-th/9803008.

\bibitem{10r} NA50 Collaboration: L. Ramello at Quark Matter '97, Tsukuba,
Japan,
December 1997; M. Gonin at Probes of Dense Matter in Ultrarelativistic Heavy
Ion Collisions,
Seattle, USA, May 1998.

\bibitem{11r} D. Kharzeev, M. Nardi and H. Satz, preprint BI-TP 97/33
(hep-ph/9707308); M. Nardi
and H. Satz, preprint BI-TP 98/10 (hep-ph/9805247).

\bibitem{12r} T. Matsui, presented at Quark Matter '97, {\it ibid}.

\bibitem{11br} S. Frankel and W. Frati, hep-ph/9710532; R. C. Hwa, J.
Pi\v{s}\'{u}t and
N. Pi\v{s}\'{u}tov\'a, Phys.
Rev. {\bf C56}, 432 (1997); {\bf C58}, 434 (1998); H. Sorge, E.
Shuryak and I.
Zahed, Phys. Rev. Lett. {\bf 79}, 2775 (1997).

\bibitem{13r} A. Capella, J. A. Casado, C. Pajares, A. V. Ramallo and J.
Tr$\hat{{\mbox
a}}$n Thanh
V$\hat{{\mbox a}}$n,
Phys. Lett. {\bf B206}, 354 (1988); A. Capella, C. Merino,
J. Tr$\hat{{\mbox
a}}$n Thanh
V$\hat{{\mbox a}}$n, C. Pajares and A. V. Ramallo,
Phys. Lett. {\bf B243},
144 (1990).

\bibitem{14r} C. W. de Jager, H. de Vries and C. de Vries, Atomic Data and
Nuclear Data Tables
{\bf 14}, 479 (1974).

\bibitem{15r} P. Koch, U. Heinz and J. 
Pi\v{s}\'{u}t, Phys. Lett. {\bf B243}, 149
(1990).

\bibitem{16r} A. Capella, Phys. Lett. {\bf B364}, 175 (1995); A. Capella,
A. B. Kaidalov, A. Kouider Akil, C. Merino and J. Tr$\hat{{\mbox
a}}$n Thanh
V$\hat{{\mbox a}}$n, Z. Phys.
{\bf C70}, 507 (1996).

\bibitem{17r} S. J. Brodsky and A. H. Mueller, Phys. Lett. {\bf B206}, 685
(1988).

\bibitem{18r} NA38 Collaboration: C. Baglin {\it et al.}, Phys. Lett. {\bf
B251}, 472 (1990).

\bibitem{19r} A. Bialas, in Proceedings of the XIIIth International Symposium
on Multiparticle
Dynamics, edited by W. Kittel, W. Metzger and A. Stergiou (World Scientific,
Singapore, 1983), p.
328.

\bibitem{21r} For a review see A. Capella, U. P. Sukhatme
C.-I. Tan and
J. Tr$\hat{{\mbox
a}}$n Thanh
V$\hat{{\mbox a}}$n, Phys. Rep. {\bf 236},
225 (1994).

\bibitem{20r} NA49 Collaboration: presented by G. Roland at Quark Matter '97,
{\it ibid}.

\bibitem{20br} NA35 Collaboration: T. Alber {\it et al.}, preprint
IKF-HENPG/6-94 (July 1997); M.
Ga\'zdzicki and D. R\"ohrich, Z. Phys. {\bf C65}, 215 (1995).

\bibitem{22r} A. Bialas, M. Bleszy\'nski and W. Czyz, Nucl. Phys. {\bf B111},
461 (1976).

\bibitem{24r} F. Bellaiche (NA50 Collaboration), Ph.D. Thesis, Universit\'e de
Lyon, France, 1997.

\bibitem{26r} A. Krzywicki, J. Engels, B. Petersson and U. P. Sukhatme,
Phys. Lett. {\bf 85B}, 407 (1979); S.
Gavin and M.
Gyulassy, Phys. Lett. {\bf B214}, 241 (1988); J.-P. Blaizot and J.-Y.
Ollitrault, Phys. Lett. {\bf
B217}, 392 (1989); S. Gupta and H. Satz, Phys. Lett. {\bf B283}, 439 (1992); S.
Gavin and R. Vogt,
preprint CU-TP-791 (hep-ph/9610432); see also Ref. \protect{\cite{2r}}.

\bibitem{27r} D. Kharzeev, M. Nardi and H. Satz, Phys. Lett. {\bf B405}, 14
(1997).

\bibitem{28br} NA38 Collaboration: C. Baglin {\it et al.}, Phys. Lett. {\bf
B262}, 362 (1991).

\bibitem{28bbr} R. Mandry (NA38 Collaboration), Ph.D. Thesis, Universit\'e
Claude Bernard,
Clermont-Ferrand, France, 1993.

\bibitem{gh} C. Gerschel and J. H\"ufner, hep-ph/9802245.

\bibitem{28r} NA3 Collaboration: J. Badier {\it et al.}, Z. Phys. {\bf C20},
101 (1983).

\bibitem{28pbr} NA38 Collaboration: M. C. Abreu {\it et al.}, Phys. Lett.
{\bf B423}, 207 (1998).

\bibitem{30r} NA50 Collaboration: presented by A. Romana at the XXXIIIrd
Rencontres de Moriond,
Les Arcs, France, March 1998.

\bibitem{31r} A. Capella, C. Merino and J. Tr$\hat{{\mbox a}}$n Thanh
V$\hat{{\mbox a}}$n, Phys.
Lett. {\bf B265}, 415 (1991).

\bibitem{32r} M. A. Braun, C. Pajares, C. A. Salgado, N. Armesto and A.
Capella, Nucl. Phys. {\bf B509}, 357 (1998).

\end{thebibliography}
\end{document}